\documentclass[aps, pra, showpacs, superscriptaddress, 10pt, twocolumn]{revtex4-1}

\usepackage{amsfonts,amssymb,amsmath,amsthm,longtable,array}
\usepackage{graphics,graphicx,color}

\usepackage{epstopdf,epsfig}
\newcommand{\ket}[1]{ | #1 \rangle}
\newcommand{\Ket}[1]{ | #1 \rangle}

\renewcommand{\>}{\rangle}
\newcommand{\be}{\begin{equation}}
\newcommand{\ee}{\end{equation}}
\newcommand{\ben}{\begin{eqnarray}}
\newcommand{\een}{\end{eqnarray}}

\usepackage{makecell}
\newcolumntype{x}[1]{>{\centering\arraybackslash}p{#1}}

\usepackage{tikz}
\usetikzlibrary[positioning]
\usetikzlibrary{patterns}
\newcommand{\daywidth}{1.02 cm}

\begin{document}
	
	\title{$k$-uniform mixed states}
	
	\author{Waldemar K{\l}obus}
	\affiliation{Institute of Theoretical Physics and Astrophysics, Faculty of Mathematics, Physics and Informatics, University of Gda\'nsk, 80-308 Gda\'nsk, Poland}
	
		\author{Adam Burchardt}
	\affiliation{Institute of Physics, Jagiellonian University, 30-348 Krak\'ow, Poland}
	
	\author{Adrian Ko{\l}odziejski}
	\affiliation{Institute of Theoretical Physics and Astrophysics, Faculty of Mathematics, Physics and Informatics, University of Gda\'nsk, 80-308 Gda\'nsk, Poland}
	
    \author{Mahasweta~Pandit}
	\affiliation{Institute of Theoretical Physics and Astrophysics, Faculty of Mathematics, Physics and Informatics, University of Gda\'nsk, 80-308 Gda\'nsk, Poland}
	
	\author{Tam\'as V\'ertesi}
    \affiliation{MTA Atomki Lend\"ulet Quantum Correlations Research Group, Institute for Nuclear Research, Hungarian Academy of Sciences, H-4001 Debrecen, P.O. Box 51, Hungary}

	\author{Karol \.Zyczkowski}
	\affiliation{Institute of Physics, Jagiellonian University, 30-348 Krak\'ow, Poland}
	\affiliation{Center for Theoretical Physics, Polish Academy of Sciences, 02-668 Warszawa, Poland}	
	\affiliation{National Quantum Information Centre in Gda\'nsk, 81-824 Sopot, Poland}
	
	\author{Wies{\l}aw Laskowski}
	\affiliation{Institute of Theoretical Physics and Astrophysics, Faculty of Mathematics, Physics and Informatics, University of Gda\'nsk, 80-308 Gda\'nsk, Poland}
	
	\affiliation{International Centre for Theory of Quantum Technologies, University of Gda\'nsk, 80-308 Gda\'nsk, Poland}

\begin{abstract}
We investigate the maximum purity that can be achieved by $k$-uniform mixed states of $N$ parties. Such $N$-party states have the property that all their $k$-party reduced states are maximally mixed. A scheme to construct explicitly $k$-uniform states using a set of specific $N$-qubit Pauli matrices is proposed.
We provide several different examples of such states and demonstrate that in some cases the state corresponds to a particular orthogonal array. The obtained states, despite being mixed, reveal strong non-classical properties such as genuine multipartite entanglement or violation of Bell inequalities. 	\end{abstract}

\maketitle
	
\section{Introduction}

Since quantum correlations are both a basic resource in quantum information processing and a fundamental phenomenon related to foundations of quantum mechanics, their characterization becomes of great importance for practical as well as strictly theoretical reasons~\cite{gt}.
For the simplest system of two qubits, the Bell states \cite{nc} play a special role. They are also known as maximally entangled states, because they exhibit strong two-qubit quantum correlations, and at the same time their single-qubit reduced states are maximally mixed. A lot of attention has recently been paid to the identification of entangled states that generalize that concept -- the pure states of $N$-partite systems, such that tracing out
arbitrary $N-k$ subsystems, the remaining $k$ subsystems are maximally mixed (see e.g.~Refs.~\cite{gisin,Higuchi} for pioneer works). Such pure states are
called $k$-uniform. By construction the integer
number $k$ cannot exceed $N/2$ and the states for $k = \lfloor N/2 \rfloor$ are called as \textit{absolutely maximally entangled} (AME) \cite{lo}. They are a natural generalization of maximally entangled Bell states ($k=1$).

While AME states for five and six qubits have been constructed explicitly~\cite{facchi, borras, scott}, such states do not exist for systems consisting of four \cite{Higuchi} and seven qubits \cite{huber}.
Moreover, it has been shown that there exist no AME states for systems with a larger number of qubits~\cite{rains1, rains2}. Interestingly, if the local dimension is chosen to be large enough, AME states always exist~\cite{cui}. For example, it has been proven that there exist AME states for three and four qu$d$its, for every prime $d > 2$ \cite{karol}. A necessary condition \cite{scott,hein} for the existence of $N$-partite AME state of arbitrary dimension is given by
\begin{equation}
N \leq
\begin{cases}
2(d^{2} -1) \quad & n \quad \text{even} \\
2d(d+1) -1 \quad & n \quad \text{odd}
\end{cases}
\end{equation}

Since for many cases one cannot construct pure $k$-uniform states, one can ask a question -- {\em what is the highest possible purity of a $k$-uniform state for a given number of parties $N$?}

In this paper we address the problem of finding $k$-uniform states with the highest possible purity for which the corresponding pure AME states do not exist.
We begin with reformulation of $k$-uniformity of states with the use of correlation tensor, then we proceed with describing the method of explicit construction of $k$-uniform states using $N$-qubit Pauli operators. Next we describe a relation between the presented construction and the notion of orthogonal arrays. In the following, we give specific examples of $k$-uniform $N$-qubit states, which also are numerically proven to be of the highest purity with respect to given values $k$ and $N$. After remarking on the properties of the $k$-uniform states with regards to entanglement and quantum Fisher information, we present an example of a specific quantum circuit which enables generating of the respective $k$-uniform state. We then briefly mention the results for $k$-uniform qu\textit{d}it states with higher dimensionality of subsystems, after which we summarize with conclusions.

\section{Correlations of $k$-uniform states}

An arbitrary state of $N$ qubits can be represented as:
\begin{equation}\label{roo}
\rho = \frac{1}{2^N} \sum_{\mu_1, \dots, \mu_N=0}^3 T_{\mu_1 \dots \mu_N} \, \sigma_{\mu_1} \otimes \dots \otimes \sigma_{\mu_N},
\end{equation}
where $\sigma_\mu$ are Pauli matrices and
$T_{\mu_1 \dots \mu_N} = \mathrm{Tr}(\rho \, \sigma_{\mu_1} \otimes \dots \otimes \sigma_{\mu_N})$ are real coefficients called correlation tensor elements which we will call simply correlations.

Let us now define {\em a length of correlations} among $r$ subsystems
\begin{equation}
M_r(\rho) = \sum_{\pi} \sum_{i_1,i_2, ..., i_r=1}^3 T_{\pi(i_1i_2...i_r)}^2,
\end{equation}
where $\pi(i_1 \dots i_r)$ stands for all permutations of $r$ non-zero indices on $N$ positions.
For a $k$-uniform state of $N$ particles $\rho_N^k$ we have
\ben\label{trzy}
M_r(\rho_N^k)= 0,
\een
for all $1 \leq r \leq k$. In other words, $k$-uniform states do not have any $k$-partite correlations, as well as correlations between smaller number of parties, i.e.  $T_{\pi(i_1i_2...i_r)} =0$ for $r\leq k$.

With this notation, the purity of a given $N$-qubit state is given by
\begin{equation}
{\rm Tr} \rho^2 =  \frac{1}{2^N}\sum_{i_1,i_2, ..., i_N=0}^3 T_{i_1i_2...i_N}^2 = \frac{1}{2^N} \left(1 + \sum_{r=1}^N M_r(\rho)\right).
\end{equation}
Furthermore, because of equation \eqref{trzy}, the sum can be reduced only to the last $N-k-1$ elements
\begin{equation}
{\rm Tr} (\rho_N^k)^2 =  \frac{1}{2^N}\left(1 + \sum_{r=k+1}^N M_r(\rho)\right).
\end{equation}

For a given purity, the total length of correlations $\sum_{r=1}^N M_r (\rho) = 2^N {\rm Tr} \rho^2 -1$ is fixed and state independent. The absence of correlation for $r \leq k$ results in the fact that all available correlations occur between a large number of qubits ($r>k$). This, combined with a relatively high purity, can manifest strong non-classical properties, for instance the genuine multipartite entanglement.

\section{States from generators}\label{sfg}

Below we present a scheme for constructing $k$-uniform states from particular sets of $N$-qubit Pauli matrices. These building blocks resemble the generators as used within the framework of stabilizer formalism \cite{stabile1,stabile2}. 
For further convenience, if not stated otherwise, we will use the simplified notation for multi-qubit Pauli operators as
\ben
\sigma_{0} \otimes \sigma_{1} \otimes \sigma_{2} \otimes \sigma_{3} \otimes ... \equiv \openone
 X Y Z ... .
 \label{6}
\een

Let us now suppose that there exists a set of $N$-qubit Pauli operators
\ben
\mathcal{G} = \{ G_1, ..., G_m \},
\label{G}
\een
such that these operators have the following properties:
\begin{itemize}
	\item[(1)] mutual commutation:  $[ G_i, G_j ] = 0$ for all $i,j$;
	\item[(2)] independence: $G^{i_1}_1 \dots G^{i_m}_m \sim \openone$  only for $i_1 =  \dots =  i_m = 0$ with $i_j=\{0,1\}$;
	\item[(3)] $k$-uniformity: $G^{i_1}_1 \dots G^{i_m}_m$  ($i_j=\{0,1\}$) results in $N$-qubit Pauli operator (\ref{6}) containing the identity operators on at most $N-k-1$ positions.
\end{itemize}
The last property distinguishes our approach from the standard stabilizer formalism (see e.g. \cite{Plenio}). In literature, $m$ is called the rank of the stabilizer group. Stabilizer groups with $m = N$ are called {\em full-rank}, whereas stabilizer groups with $m < N$ {\em rank-deficient}. The elements of such a set $\mathcal{G}$ will be called generators. We can use them to generate a $k$-uniform state by summing all possible products of the elements from $\mathcal{G}$
\begin{equation}\label{rofromgen}
\rho=\frac{1}{2^N} \sum_{j_1, ... j_m=0}^1   G^{j_1}_1 ... G^{j_m}_m.
\end{equation}
The above construction leads to a valid physical state by virtue of the following argument.

Consider a set of $m$ mutually commuting $N$-qubit Pauli operators $\mathcal{G} = \{ G_1, ..., G_m \}$. Let us rewrite the state \eqref{rofromgen} into the form
\ben
\rho=\frac{1}{2^N} (\openone + G_1)(\openone + G_2) \ldots (\openone + G_m).
\een
Therefore, we see that the eigenvalues of $\rho$ can be written in the form
\ben
\lambda^i = \frac{1}{2^N} (1+\lambda^i_1)(1+\lambda^i_2) \ldots (1+\lambda^i_m),
\een
where $\lambda^i_j = \pm 1$ is the $i$-th eigenvalue of the $j$-th generator in common eigenbasis of mutually commuting operators from the set $\mathcal{G}$. Note that from \eqref{rofromgen} we have $\textrm{Tr}\rho =1$, while $\lambda^i$ are either 0 or $2^{m-N}$, hence $\rho$ constitutes a physical state with exactly $2^{N-m}$ nonzero eigenvalues. Naturally, the case $m=N$ corresponds to a pure state with exactly one eigenvalue equal to 1.

Now, the state \eqref{rofromgen} has $2^m$ non-vanishing correlations equal to $\pm 1$ and its purity can be calculated simply as
\ben
{\rm Tr} \rho^2 = \frac{1}{2^N} 2^m =2^{m-N}.
\een

Note that the larger the set $\mathcal{G}$, the higher the purity of the outcoming state is.
We observe that the problem of constructing $k$-uniform states is therefore directly related to the problem of finding the largest possible set of generators $\mathcal{G}$.
Consequently, in the case of $k = \left\lfloor \frac{N}{2} \right\rfloor$ and $m=N$ the construction leads to an AME state with purity equal to 1.

Due to the construction method we expect to obtain $k$-uniform states of high purity. In all considered cases (up to $N=6$)
we have numerical evidence that there are no $k$-uniform states of higher purity (see Appendix \ref{num} for details).

\section{Orthogonal arrays}

In general, in order to determine $\mathcal{G}$ we have to search the full set of $4^N$ $N$-qubit operators.
However, we observe that it is possible to construct a set of generators with the help of orthogonal arrays.
Orthogonal arrays \cite{rao, Hedayat} are combinatorial arrangements, tables with entries satisfying given orthogonal properties.
An orthogonal array ${\rm OA}(r,N,l,s)$ is a table composed by $r$ rows, $N$ columns with entries taken from $0,\ldots,l-1$ in such a way that each subset of $s$ columns contains all possible combination of symbols with the same amount of repetitions. The number of such repetitions is called \textit{the index} of the OA; if $l^s =r$ orthogonal array is \emph{of index unity}.

Suppose we wish to find $\mathcal{G}$ for a $k$-uniform state of $N$ qubits. For this purpose we can use an orthogonal array ${\rm OA}(r,\textit{\textbf{N}},{\bf 4},s)$ with 4 levels (corresponding to four different Pauli matrices).
In doing so, we treat each row of OA as a string of indices $a_1 \dots a_N$ ($a_i \in \{0,1,2,3\}$), which defines the specific $N$-qubit Pauli operator $A_1 \dots A_N$ ($A_i \in \{\openone,X,Y,Z\}$) using a convention: $0 \to \openone$, $1 \to X$, $2 \to Y$ and $3 \to Z$. After performing this operation we end up with a set of $r$ operators, from which we have to choose the largest set $\mathcal{G}$  such that its elements meet the conditions (1-2) from (\ref{G}). Those conditions guarantee that desired state is physical and determine its purity. The parameter $k$ for which the property (3) from (\ref{G}) holds does not depend explicitly on the presented construction but rather on a particular example of OA. The maximal number of $\openone$'s in each row of OA equals to $s-1$, which may suggest $(N-s)$-uniformity of obtained state. In condition (3), however, we require that the number of $\openone$'s is limited not only for generators but also for all elements of the form $G^{i_1}_1 \dots G^{i_m}_m$. In some of presented examples (see Secs. \ref{D}, \ref{F}, \ref{H}) the number of $\openone$'s is also limited by $s-1$ for all such elements. Hence the desired states are indeed $(N-s)$-uniform. In other examples, however, uniformity of the desired state is slightly smaller than the prediction from the generators. Although the states obtained from OA of index unity coincide with $(N-s)$-uniform states, the precise connection has to be established. In general, the relation between uniformity $k$ and quantities $s$ and $N$ seems to be irregular.

It is well know that in the simplest case of four qubits there is no $2$-uniform pure state \cite{Higuchi}.
However, relaxing the assumption that desired state is pure, the orthogonal array {\rm OA}$(16,4,4,2)$ can be utilized to construct the mixed 4-qubit 2-uniform state. It leads to the following set of operators
\begin{eqnarray}
&0000 \to \openone \openone \openone \openone, \;\;\; 0111 \to \openone XXX,& \nonumber \\
&0222  \to \openone YYY, \;\;\; 0333  \to \openone ZZZ,& \nonumber \\
&1012  \to X\openone XY, \;\;\; 1103  \to XX\openone Z,& \nonumber \\
&1230  \to XYZ\openone, \;\;\; 1321 \to XZYX,& \nonumber \\
&2023  \to Y\openone YZ, \;\;\; 2132  \to YXZY,& \nonumber \\
&2201  \to YY\openone X, \;\;\; 2310  \to YZX\openone,& \nonumber \\
&3031  \to Z\openone ZX, \;\;\; 3120  \to ZXY\openone,& \nonumber \\
&3213  \to ZYXZ, \;\;\; 3302 \to ZZ\openone Y.& \nonumber \\
\end{eqnarray}
Within this set one can find $m=3$ operators conforming to the properties from Sec. \ref{sfg}, which constitute the set $\mathcal{G}$, e.g.:
\begin{eqnarray}
G_1 &=& \openone YYY, \nonumber \\
G_2 &=& XZYX, \\
G_3 &=& YXZY \nonumber
\end{eqnarray} and by virtue of Eq. (\ref{rofromgen}), leads to the state $\rho_4^2$ of purity $1/2$.

\section{Examples}\label{examp}

Below we present examples of $k$-uniform states with the highest possible purity for several cases of $k$ and $N$. In each case we provide generators from the set $\mathcal{G}$, which uniquely define the corresponding $k$-uniform state. All examples are summarized in Fig. 1.

\subsection{General schemes}
\label{General schemes}

When the verification of the properties (1--3) for generators in Sec.~\ref{sfg} becomes computationally demanding, in some particular cases we can employ simple schemes for construction of $k$-uniform $N$-qubit states. (i) The first method, presented in details in \cite{NOCORR1, NOCORR2}, can be implemented if other particular $(k-1)$-uniform state is known ($k-1$ is even). The method eliminates all correlations between odd number of subsystems and does not change the remaining ones.
Since $k$ is odd for even $k-1$, the $k$-partite correlations vanish and the state becomes $k$-uniform. To this end we evenly mix the original state $\rho_{N}^{k-1}$ with its `antistate':
\begin{eqnarray}
\rho_{N}^{k}=\frac{1}{2}(\rho_{N}^{k-1} + \bar{\rho}_N^{k-1}),
\end{eqnarray}
where the `antistate' $\bar{\rho}_N^{k-1}= \sigma_y^{\otimes N} {\rm conj}(\rho_N^{k-1}) \sigma_y^{\otimes N} $ and ${\rm conj}(.)$ denotes the complex conjugation. (ii) We can also obtain $k$-uniform states by tracing out any of subsystems of $N$-qubit $k$-uniform state. It leads to the $k$-uniform $(N-1)$-qubit state. In both methods the purity of the resulting state is reduced by a half. These methods, however, do not guarantee that the obtained states are of the highest possible purity.

\subsection{$N$ arbitrary, $k=1$}
The  $1$-uniform pure state is the $N$-qubit GHZ state $| \textrm{GHZ}\> = 1/\sqrt2 (| 0...0\>+| 1...1\>)$, for which $m=N$ generators are
\begin{eqnarray}
&G_1 = ZX \cdots XX,\;\;\; G_2 = XZ \cdots XX,&  \\
&\cdots& \nonumber \\
&G_{N-1} = XX \cdots ZX,\;\;\; G_{N} = XX \cdots XZ.&
\end{eqnarray}

\subsection{$N$ arbitrary, $k=N-1$}
\label{fact1}
For $k=N-1$ only $N$-partite correlations are possible, hence the generators cannot have identity operator $\openone$ on any position.
For $N$ odd, the set $\mathcal{G}$ consists of only one  generator $(m=1)$:
\begin{eqnarray}
G_1 = Z \cdots Z,
\end{eqnarray}
while for $N$ even, the set $\mathcal{G}$ consists of two generators $m=2$:
\begin{eqnarray}
G_1 = X \cdots X,\;\;\; G_2 = Z \cdots Z.
\end{eqnarray}
For even $N$ the states can be written in the following form:
\begin{eqnarray}
\rho_{N} = \frac{1}{2^N} \left(\openone^{\otimes N} + (-1)^{N/2} \sum_{j=1}^3 \sigma_j^{\otimes N}\right)
\end{eqnarray}
and are known as the generalized bound entangled Smolin states \cite{smolin1, smolin2}. They are a useful quantum resource for multiparty communication schemes and were experimentally demonstrated \cite{smolin3}.

\subsection{$N=4$, $k=2$} \label{D}
Since in this case a pure AME state does not exist \cite{Higuchi}, we cannot have four generators, so the set $\mathcal{G}$ consists of $m=3$ elements:
\begin{eqnarray}
&G_1 = XXXX,\;\;\; G_2 = YYYY,&  \\
&G_3 = \openone XYZ.& \nonumber
\end{eqnarray}

\noindent
The above construction yields the symmetric mixture of two pure states:
\begin{eqnarray}
\label{42a}
\Ket{\varphi_1} =\dfrac{1}{\sqrt{2}}    \Big( \Ket{\phi_1} &+&\Ket{\phi_2}\Big),\\
\Ket{\varphi_2} =\dfrac{\sigma_{x}^{\otimes 4}}{\sqrt{2}}       \Big(  \Ket{\phi_1}& -& \Ket{\phi_2}\Big),\nonumber
\end{eqnarray}
where
\begin{eqnarray*}
\Ket{\phi_1} = \dfrac{1}{2}
 \Big( \Ket{0010}+ \Ket{1110} +i \Ket{1000} -i\Ket{1001}\Big), \\
 \Ket{\phi_2} = \dfrac{1}{2}
  \Big( \Ket{1111} - \Ket{0011} +i \Ket{0100} +i \Ket{0101} \Big),
\end{eqnarray*}
\noindent
and $\sigma_{x}^{\otimes 4}$ is a flip operation on all particles.
Notice that each of states given in equation \eqref{42a} is almost $2$-uniform. More accurately, 4 out of ${4 \choose 2}$ its reductions to 2 qubits are maximally mixed. The remaining reductions are given in standard basis by:
\[\dfrac{1}{4}
\left(\begin{array}{cccc}
 1 & -1 & 0 & 0\\
    -1       & 1 & 0& 0 \\
  0    & 0 & 1 & 1 \\
    0      & 0& 1& 1
\end{array}\right),
\text{     and     }
\dfrac{1}{4}
\left(\begin{array}{cccc}
 1 & 1 & 0 & 0 \\
    1       & 1 & 0& 0 \\
  0    & 0 & 1 & -1 \\
    0      & 0& -1& 1
\end{array}\right)
\]
\noindent
for $\Ket{\varphi_1}$ and $\Ket{\varphi_2}$, respectively. Observe that the sum of those matrices is proportional to $\openone$, which is relevant to the fact that the mixture of $\Ket{\varphi_1}$ and $\Ket{\varphi_2}$ is 2-uniform.


\subsection{$N=5$, $k=2$}
The 5-qubit pure AME state is described by $m=5$ generators:
\begin{eqnarray}
&G_1 = \openone XYXY,\;\;\; G_2 = \openone ZXX \openone,& \nonumber  \\
&G_3 = XYY \openone Z,\;\;\; G_4 = XZYZY,&  \\
&G_5 = ZXZ \openone X.& \nonumber
\end{eqnarray}

\noindent
The explicit formula of the state is:
\begin{eqnarray*}
\dfrac{1}{\sqrt{8}}\Big( \Ket{01111}+ \Ket{10011}  + \Ket{10101} + \Ket{11100} &-& \\
\big( \Ket{00000}+ \Ket{00110} + \Ket{01001} + \Ket{11010} \big) \Big),&&
\end{eqnarray*}
and is equivalent to the AME(5,2) state constructed via the link with quantum error correction codes \cite{QECC}.

\subsection{$N=5$, $k=3$} \label{F}
The 5-qubit 3-uniform mixed state can be obtained from OA$(16,5,4,2)$, which leads to the following $m=4$ generators:
\begin{eqnarray}
&G_1 = \openone XXXX,\;\;\; G_2 = \openone YYYY,& \nonumber  \\
&G_3 = X \openone XYZ,\;\;\; G_4 = Y \openone YZX.&
\end{eqnarray}
The corresponding state is of the form \eqref{42a}, with
\begin{eqnarray*}
	\Ket{\phi_1} =\dfrac{1}{2}
	\Big(\Ket{001 0 1} + \Ket{01 0 1 0} +i  \Ket{001 1 0} +i\Ket{01 0 0 1}  \Big), \\
	\Ket{\phi_2} = \dfrac{1}{2}
	\Big( \Ket{10 0 00}+ \Ket{1 1 1 1 1}  -i\Ket{10 0 1 1} -i \Ket{11 1 0 0} \Big).
\end{eqnarray*}
\noindent
and has purity $1/2$.

An interesting property of the state is the fact that it contains only four-qubit correlations.
Nevertheless, the state is genuinely five-qubit entangled.

\subsection{$N=6$, $k=2$}
The 6-qubit 2-uniform pure state can be described by $m=6$ generators:
\begin{eqnarray}
&G_1 = XXYYZZ,\;\;\; G_2 = XXZZYY,& \nonumber \\
&G_3 = XZZXXZ,\;\;\; G_4 = XYYX \openone Z,& \\
&G_5 = YX \openone Z XY,\;\;\; G_6 = YYYY \openone \openone.& \nonumber
\end{eqnarray}
and is equivalent to the state presented in \cite{GZ}.

\subsection{$N=6$, $k=3$} \label{H}
The 6-qubit AME state can be obtained from OA$(64,6,4,3)$, which leads to the following $m=6$ generators
\begin{eqnarray}
&G_1 =  \openone \openone ZZZZ,\;\;\; G_2 = \openone XYZ \openone X,& \nonumber \\
&G_3 = \openone ZXY \openone Z,\;\;\; G_4 = XYZ \openone Z \openone,&  \\
&G_5 = Z \openone Z \openone XY,\;\;\; G_6 = ZYYZZY.& \nonumber
\end{eqnarray}
Formula (\ref{rofromgen}) gives a pure AME(6,2) state
\begin{eqnarray*}
\Ket{\phi} =&\dfrac{1}{4}
\Big( \Ket{000110}+ \Ket{011100}  +\Ket{100000} + \Ket{111010} \\
&-\Ket{001001}- \Ket{010011}  -\Ket{101111} - \Ket{110101} \\
&+ i\Ket{000101}+ i\Ket{010000}  +i\Ket{101100} + i\Ket{111001} \\
&-i\Ket{001010}- i\Ket{011111}  -i\Ket{100011} - i\Ket{110110} \Big).
\end{eqnarray*}
equivalent to the one found in \cite{karol}.

\subsection{$N=6$, $k=4$}
The  4-uniform 6-qubit  mixed state can be described by $m=3$ generators:
\begin{eqnarray}
&G_1 =  \openone XXXXX,\;\;\; G_2 = Y \openone YYYY,& \nonumber \\
&G_3 = ZZ \openone XYZ,& \nonumber
\end{eqnarray}
and has purity $1/8$.

\subsection{$N=7$, $k=2$}
The   2-uniform 7-qubit pure state can be described by $m=7$ generators:
\begin{eqnarray}
&G_1 =  \openone XXYYZZ,\;\;\; G_2 = \openone Z \openone \openone Z \openone Z,& \nonumber \\
&G_3 = XXZXZZX,\;\;\; G_4 = YYZYZZY,& \nonumber \\
&G_5 = \openone \openone ZXYYX,\;\;\; G_6 = \openone YY \openone \openone YY,& \\
&G_7 = Z \openone Z \openone Z \openone Z,& \nonumber
\end{eqnarray}
which results in the state of the form \cite{GZ}:
\begin{eqnarray*}
\Ket{\phi} =&\dfrac{1}{\sqrt{8}}
\Big( \Ket{0000000}+ \Ket{01 1 0011}  +\Ket{101 1 010} + \Ket{11 0 1 0 0 1} \\
&-\Ket{1111111}- \Ket{01111 00}  -\Ket{1 0 1 0 1 01} - \Ket{1100110} \Big).
\end{eqnarray*}

\subsection{$N=7$, $k=3$}
Since in this case a pure AME state does not exist, we cannot specify 7 generators. Here, however, we can employ the scheme for eliminating all the correlations of the rank given by even number. Therefore, using the above 7-qubit 2-uniform state $\rho^2_7$, we can construct 7-qubit 3-uniform mixed state $\rho^3_7$, which is described by $m=6$ generators:
\begin{eqnarray}
&G_1 = Y\openone YXZXZ,\;\;\; G_2 = \openone XXYYZZ,& \nonumber \\
&G_3 = ZXYYXZ\openone,\;\;\; G_4 = ZZ\openone YXXY,& \\
&G_5 = YY\openone Y\openone \openone Y,\;\;\; G_6 = ZXYZ\openone YX.& \nonumber
\end{eqnarray}

\subsection{$N=7$, $k=5$}
The 7-qubit 5-uniform mixed state can be obtained from the following $m=3$ generators:
\begin{eqnarray}
&G_1 = \openone XXXXXX,\;\;\; G_2 = X \openone X YYYY,& \nonumber \\
&G_3 = YYY \openone XYZ.& \nonumber
\end{eqnarray}
The purity of the state is 1/16.

\subsection{$N>7$, $k=N-2$}
\label{fact2}
For $k=N-2$ only $(N-1)$-partite correlations are possible, hence the generators have the identity operator $\openone$ on at most one position. In either case, at least two generators can be found, for $N$ odd:
\begin{eqnarray}
G_1 =  \openone X \cdots X,\;\;\; G_2 =  \openone  Y\cdots Y,
\end{eqnarray}
while for $N$ even:
\begin{eqnarray}
G_1 =  \openone X \cdots X,\;\;\; G_2 = X \openone  Y\cdots Y,
\end{eqnarray}
leading to purity $1/2^{N-2}$.

\subsection{$N=9$, $k=5$}
The 9-qubit 5-uniform mixed state can be obtained from OA$(32,9,4,2)$, which leads to the following $m=4$ generators:
\begin{eqnarray}
&G_1 = XXXXXXXX\openone,\;\;\; G_2 = YYYYYYYY\openone,& \nonumber \\
&G_3 = \openone XYZ \openone XYZX,\;\;\; G_4 = \openone \openone ZZYYXX \openone.
\end{eqnarray}
The purity of the state is $1/32$.

\subsection{$N=12$, $k=5$}
From OA(4096,12,4,5) we can isolate the following set of $m=6$ generators:
\begin{eqnarray}
G_1 &=& XYY \openone ZZX \openone \openone \openone \openone \openone, \nonumber \\
G_2 &=& YZZ \openone XXY \openone \openone \openone \openone \openone, \nonumber \\
G_3 &=& \openone XYY \openone ZZX \openone \openone \openone \openone, \\
G_4 &=& \openone YZZ \openone XXY \openone \openone \openone \openone, \nonumber \\
G_5 &=& XXXXXXXXXXXX, \nonumber \\
G_6 &=& YYYYYYYYYYYY, \nonumber
\end{eqnarray}
which leads to the state of purity 1/64.
\begin{figure}
\begin{tikzpicture}[x=\daywidth, y=-1cm, node distance=0 cm,outer sep = 0pt]
\tikzstyle{day}=[draw, rectangle,  minimum height=1cm, minimum width=\daywidth, fill=white!15,anchor=south west]
\tikzstyle{day2}=[draw, rectangle,  minimum height=1cm, minimum width=1.0 cm, fill=white!20,anchor=south east]
\tikzstyle{hour}=[draw, rectangle, minimum height=1 cm, minimum width=1.0 cm, fill=white!30,anchor=north east]
\tikzstyle{1hour}=[draw, rectangle, minimum height=1 cm, minimum width=\daywidth, fill=white!30,anchor=north west]
\tikzstyle{Planche0}=[1hour,fill=red!40]
\tikzstyle{Planche1}=[1hour,fill=blue!60]
\tikzstyle{Planche2}=[1hour,fill=blue!50]
\tikzstyle{Planche3}=[1hour,fill=blue!40]
\tikzstyle{Planche4}=[1hour,fill=blue!30]
\tikzstyle{Planche5}=[1hour,fill=blue!20]
\tikzstyle{Planche6}=[1hour,fill=blue!10]
\tikzstyle{Planche7}=[1hour,fill=blue!5]
\tikzstyle{Planche8}=[1hour,fill=blue!2]
\tikzstyle{Ang2h}=[2hours,fill=green!20]
\tikzstyle{Phys2h}=[2hours,fill=red!20]
\tikzstyle{Math2h}=[2hours,fill=blue!20]
\tikzstyle{TIPE2h}=[2hours,fill=blue!10]
\tikzstyle{TP2h}=[2hours, pattern=north east lines, pattern color=magenta]
\tikzstyle{G3h}=[3hours, pattern=north west lines, pattern color=magenta!60!white]
\tikzstyle{Planche}=[1hour,fill=white]
\node[day] (lundi) at (1,8) {4};
\node[day] (mardi) [right = of lundi] {5};
\node[day] (mercredi) [right = of mardi] {6};
\node[day] (jeudi) [right = of mercredi] {7};
\node[day] (vendredi) [right = of jeudi] {8};
\node[day] (nine) [right = of vendredi] {9};
\node[hour] (8-9) at (1,8) {1};
\node[hour] (9-10) [below = of 8-9] {2};
\node[hour] (10-11) [below= of 9-10] {3};
\node[hour] (11-12) [below = of 10-11] {4};
\node[hour] (12-13) [below  = of 11-12] {5};
\node[hour] (13-14) [below = of 12-13] {6};
\node[hour] (14-15) [below = of 13-14] {7};
\node[hour] (15-16) [below = of 14-15] {8};
\node[Planche0] at (1,8) {$1$};
\node[Planche1] at (1,9) {$\dfrac{1}{2}$};
\node[Planche2] at (1,10) {$\dfrac{1}{2^2}$};

\node[Planche0] at (2,8) {$1$};
\node[Planche0] at (2,9) {$1$};
\node[Planche1] at (2,10) {$\dfrac{1}{2}$};
\node[Planche4] at (2,11) {$\dfrac{1}{2^4}$};

\node[Planche0] at (3,8) {$1$};
\node[Planche0] at (3,9) {$1$};
\node[Planche0] at (3,10) {$1$};
\node[Planche3] at (3,11) {$\dfrac{1}{2^3}$};
\node[Planche4] at (3,12) {$\dfrac{1}{2^4}$};

\node[Planche0] at (4,8) {$1$};
\node[Planche0] at (4,9) {$1$};
\node[Planche1] at (4,10) {$\dfrac{1}{2}$};
\node[Planche4] at (4,12) {$\dfrac{1}{2^4}$};
\node[Planche6] at (4,13) {$\dfrac{1}{2^6}$};

\node[Planche0] at (5,8) {$1$};
\node[Planche6] at (5,13) {$\dfrac{1}{2^6}$};
\node[Planche6] at (5,14) {$\dfrac{1}{2^6}$};

\node[Planche0] at (6,8) {$1$};
\node[Planche5] at (6,12) {$\dfrac{1}{2^5}$};
\node[Planche7] at (6,14) {$\dfrac{1}{2^7}$};
\node[Planche8] at (6,15) {$\dfrac{1}{2^8}$};

\node[day2] at (1,8) {};
\draw (1,8)--(0, 7);
\node at (0.75,7.36) {$N$};
\node at (0.31,7.7) {$k$};

\draw [>=stealth,red,<-,thick] (1.5,10.15)--(1.5, 9.85);
\draw [>=stealth,red,<-,thick] (3.5,12.15)--(3.5, 11.85);
\draw [>=stealth,red,<-,thick] (4.5,10.15)--(4.5, 9.85);

\draw [>=stealth,red,<-,thick] (1.75,9.5)--(2.25, 9.5);
\draw [>=stealth,red,<-,thick] (1.75,10.5)--(2.25, 10.5);
\draw [>=stealth,red,<-,thick] (2.75,10.5)--(3.25, 10.5);
\draw [>=stealth,red,<-,thick] (2.75,11.5)--(3.25, 11.5);
\end{tikzpicture}
\caption{\label{tab1} The purity of the states presented in Section \ref{examp}.
	States that can be obtained by one of two general procedures described in Section \ref{General schemes} are indicated by horizontal (i) and vertical (ii) arrows.}
\end{figure}

\section{Genuine multipartite entanglement}

We also focus on the genuine $N$-partite entanglement for the considered $N$-qubit states.
We evaluate entanglement monotone $W$ as proposed in Ref.~\cite{GME2011} for the states with $N \leq 6$. Nonzero value of $W$ indicates genuine multipartite entanglement for the considered state.
 We find that most of the studied states exhibit genuine multipartite entanglement. The values of $W$ are presented in Table~\ref{tab}. For 7-qubit $k$-uniform states we derive witnesses using the method designed for the stabilizer states shown in \cite{stabile2, stabile1}. They have a form: $\mathcal{W}_7^k = \alpha_7^k \openone - \rho_7^k$ with $\alpha_7^1 = \alpha_7^2 = 1/2, \alpha_7^3=1/4$ and prove genuine multipartite entanglement for the considered states: $\rho_7^1, \rho_7^2$ and $\rho_7^3$.

\section{Fisher information}

Let us consider a $N$-qubit Hamiltonian that allows observers to perform a different evolution on each particle. The local evolutions are generated by the operators $\sigma_{\vec n}^{(j)} = \vec n^{(j)} \cdot \vec \sigma^{(j)}$ ($j=1, \dots, N$). Such a Hamiltonian takes the form
\begin{equation}
\mathcal{H} = \frac{1}{2}\sum_j \sigma_{\vec n_j}^{(j)},
\end{equation}
and is a generalization of a standard collective Hamiltonian for which $\sigma_{\vec n_j}^{(j)} = \sigma_{\vec n}^{(j)}$ for all $j$.

For pure states, the quantum Fisher information \cite{Caves} can be easily calculated as the variance of the Hamiltonian, $F(\rho, \mathcal{H}) = 4{\rm Tr}((\Delta \mathcal{H})^2 \rho)$. The square of the Hamiltonian is given by
\begin{equation}
\mathcal{H}^2 = \frac{N}{4} + \frac{1}{2} \sum_{i<j} \sigma_{\vec n}^{(i)}\sigma_{\vec m}^{(j)}.
\end{equation}

Therefore the quantum Fisher information can be expressed in terms of correlation tensor elements in the following way (see \cite{MARK} for comparison with a collective case):
\begin{eqnarray}
F(\rho, \mathcal{H}) &=& 4\{{\rm Tr}(\mathcal{H}^2 \rho) - ({\rm Tr}(\mathcal{H} \rho)^2\} \label{ghzgr} \\ &=&
N + 2(T_{n_1n_20...0}+ T_{n_10n_30...0} + \cdots T_{0...0n_{N-1}n_N}) \nonumber             \\ &-&(T_{n_10...0}+T_{0n_20...0}+\cdots +T_{0...0n_N})^2 \nonumber
\end{eqnarray}

Since for  $k$-uniform states (with $k\geq2$) all two- and single-qubit correlation tensor elements vanish, the quantum Fisher information,
\begin{equation}
F(\rho_N^{k\geq 2}, \mathcal{H})  = N,
\label{qfi}
\end{equation}
depends only on the number of qubits.

Note that the quantum Fisher information (\ref{qfi}) does not depend on a particular choice of the vectors $\vec n_j$. This implies that the quantum Fisher information $F_{avg}$ averaged over all directions $\vec n$  is also equal to $N$. This fact can be used to verify the presence of entanglement, because for all product states $F_{avg} < 2N/3$ \cite{Favg, Favg1}.

The situation for mixed states is more complicated. In this case Eq.~(\ref{qfi}) provides only an upper bound on the quantum Fisher information. In general it can be a function of higher order correlations. In spite of this, in several cases of mixed states we observe similar behavior as for pure states (see Tab.~\ref{tab}).

\section{Bell violation}
\label{BeelV}

We investigated considered families of $k$-uniform states (up to 7 qubits) with a numerical method based on linear programming \cite{GRUCA}. The method allows us to reveal nonclassicality even without direct knowledge
of Bell’s inequalities for the given problem. For each state we determine the minimal admixture of white noise that is necessary to destroy quantum correlations  $f_{crit}$ and the probability of violation of local realism $p_v$ for randomly sampled settings \cite{RANDOM}. The results are presented in Tab.~\ref{tab}. In all cases (except the trivial ones), $f_{crit}>0$ and we observe a conflict with local realism.

\section{Quantum circuits for $k$-uniform states}

\begin{figure}[h!]
		\includegraphics{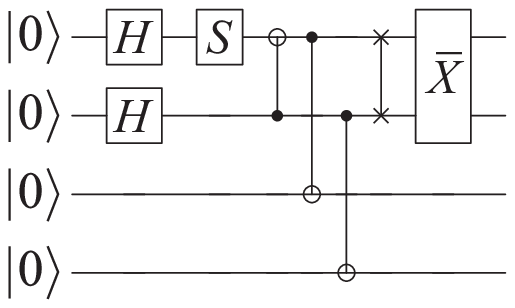}
		\caption{\label{circuit} A scheme for constructing a 2-uniform 4-qubit state. The last transformation applied at random (i.e. with probability $\frac12$) outputs an equal mixture of two pure states.}
	\end{figure}

Recently, in Ref.~\cite{qcirc} quantum circuits that generate absolutely maximally entangled states have been designed. We can employ a similar scheme in order to generate mixed $k$-uniform states. As an example, in the following we present a quantum circuit which results in generating 2-uniform 4-qubit state. The circuit is presented in Fig.~\ref{circuit} and consists of: the Hadamard operations ($H$), the phase gate ($S$), nonlocal CNOT and SWAP operations defined in a standard way as in Ref.~\cite{nc}:
		\begin{eqnarray}
	&{\rm H} = \frac{1}{\sqrt{2}}\left(\begin{array}{cc}
	1&1\\1&-1
	\end{array}\right), ~~~
	{\rm S} = \left(\begin{array}{cc}
	1&0\\0&i
	\end{array}\right),&\nonumber \\
	&{\rm CNOT} = \left(\begin{array}{cccc}
	1&0&0&0\\
	0&1&0&0\\
	0&0&0&1\\
	0&0&1&0
		\end{array}\right),~~~
			{\rm SWAP} = \left(\begin{array}{cccc}
		1&0&0&0\\
		0&0&1&0\\
		0&1&0&0\\
		0&0&0&1
		\end{array}\right).&\nonumber
		\end{eqnarray}
Since the output of quantum circuits are pure states, in order to obtain a mixed state, the last gate is applied at random: with probability $\frac12$ we perform $\bar{X}=X \otimes X$ transformation and with probability $\frac12$ we do nothing, so that the resulting state is an equal mixture of two original pure states given in Eq.~\eqref{42a}.

\section{Higher dimensional $k$-uniform states}

The scheme of generating $k$-uniform states with the use of a specific set of generators can be extended to higher dimensional systems, in which instead of $N$-qubit Pauli operators $G_i$, one uses $N$-qu\textit{d}it operators, $G^{(d)}_i$, composed of $d$-dimensional Weyl-Heisenberg matrices $S_{kl}^{(d)} = (X^{(d)})^k (Z^{(d)})^l$, where $X^{(d)}= \sum_{i=0}^d |i\rangle \langle i+1|$ and $Z^{(d)} = \sum_{i=0}^d \omega^i |i\rangle \langle i|$ with $\omega = e^{2i \pi/d}$ and $k,l=0, \ldots, d-1$ . Then the set of generators $\mathcal{G}^{(d)}$ must also conform to the same set of properties defined in Section \ref{sfg}. The resulting $k$-uniform $N$-qu$d$it state is given by
\ben
\rho=\frac{1}{d^N} \sum_{j_1,...,j_m=0}^{d-1}  (G_1^{(d)})^{j_1} ... (G_m^{(d)})^{j_m},
\een
and its purity is $d^{m-N}$. It is worth noting that if a pure $k$-uniform state does not exist, the highest purity that can be achieved is $1/d$. Already for $d=3$ this value is relatively small.

Using the above scheme one can construct so called graph states including the (1-uniform) $N$-qu$d$it GHZ-type state which is obtained from the following $m=N$ generators \cite{ghzgr}:
\begin{eqnarray}
&G_1^{(d)} = Z^{(d)}X^{(d)} \cdots X^{(d)}X^{(d)},&\nonumber\\
&G_2^{(d)} = X^{(d)}Z^{(d)} \cdots X^{(d)}X^{(d)},&  \nonumber\\
&\cdots&  \\
&G_{N-1}^{(d)} = X^{(d)}X^{(d)} \cdots Z^{(d)}X^{(d)},&\nonumber\\
&G_{N}^{(d)} = X^{(d)}X^{(d)} \cdots X^{(d)}Z^{(d)}.&\nonumber
\end{eqnarray}

Let us now ilustrate the above method with two more examples of constructing $k$-uniform qutrit states (pure and mixed). For four qutrits, as opposed to the qubit case, there exists the pure AME(4,3) state that can be determined by $m=4$ generators:
\begin{eqnarray}
&G_1^{(3)} = \openone^{(3)} Z^{(3)}Z^{(3)}(Z^{(3)})^2,& \nonumber \\
&G_2^{(3)} = \openone^{(3)} X^{(3)}X^{(3)}(X^{(3)})^2,& \\
&G_3^{(3)} = Z^{(3)} \openone^{(3)}Z^{(3)}Z^{(3)},& \nonumber \\
&G_4^{(3)} = X^{(3)} \openone^{(3)}X^{(3)}X^{(3)}.& \nonumber
\end{eqnarray}

Another example is a 2-uniform 3-qutrit mixed state defined by the following $m=2$ generators:
\begin{equation}
G_1^{(3)} = X^{(3)} X^{(3)} X^{(3)}, ~~~    G_2^{(3)}=Z^{(3)} Z^{(3)} Z^{(3)}.
\end{equation}
and can be expressed as a symmetric mixture of three pure states:
\begin{eqnarray}
\ket{\alpha_1} &=& \frac{1}{\sqrt{3}} (\ket{000} + \ket{111} +\ket{222}),\nonumber \\
\ket{\alpha_2} &=& \frac{1}{\sqrt{3}} (\ket{021} + \ket{102} +\ket{210}),\label{3q}\\
\ket{\alpha_3} &=& \frac{1}{\sqrt{3}} (\ket{012} + \ket{201} +\ket{120}).\nonumber
\end{eqnarray}
Although this state has a relatively low purity $1/3$,  it exhibits genuine multipartite entanglement ($W=1$). Note that the purity of the 2-uniform 3-qutrit state is higher than that for the corresponding qubit state (see Sec. \ref{fact1}), which is equal to $1/4$.

Finally, we used an iterative method based on semidefinite programming~\cite{SDP} to determine the maximal purity of  $k$-uniform qu$d$it states. The method is described in detail in Appendix~\ref{sdpnum}. With this algorithm, we firstly managed to reproduce all purity values up to $N=6$ parties in Table~\ref{tab}. Then we ran the algorithm for higher $d$ values.
For three parties ($N=3$, $k=2$), for $d=3$ we obtained the maximal purity equal to $1/3$ -- see Eq. \eqref{3q}, while it reads $1/4$ for $d=4$. In addition, we investigated the case ($N=4$, $k=3$, $d=3$), for which we get purity $1/9$. In this particular case the purity seems to decay with the increase of dimension.

\section{Conclusions}

We investigated the instances of $k$-uniform states of $N$ qubits, for which it is known that the corresponding absolutely maximally entangled pure states do not exist. The $k$-uniform states are distinguished by revealing the highest multipartite correlations among all quantum states of the same purity. A general scheme for finding particular sets of $N$-qubit Pauli operators, allows us to construct $k$-uniform mixed states for this system. We illustrated this method with examples of all $k$-uniform states up to 6 qubits. These states were numerically verified to be of the highest purity with respect to any given values of $k$ and $N$.

We showed that particular mixed $k$-uniform states can be constructed with the help of orthogonal arrays, but in different way from the known scheme of utilizing the notion of OA for constructing pure AME states: in the case of mixed states the key role is played by the correlation tensor elements instead of ket vectors of the pure AME state itself. We also discussed some instances of $k$-uniform states of 3- and 4-qu$d$it systems. Here, however, the dimensionality of the total system rises much faster with the number of qu$d$its making the numerical analysis ineffective for high dimensions.

\section*{Acknowledgments}

We thank Lukas Knips and Krzysztof Szczygielski for valuable discussions and Dardo Goyeneche for valuable remarks.
WK, MP and WL acknowledges the support by DFG (Germany) and NCN (Poland) within the joint funding initiative ``Beethoven2'' (2016/23/G/ST2/04273). WL acknowledge partial support by the Foundation for Polish Science (IRAP project, ICTQT, contract no. 2018/MAB/5, co-financed by EU via Smart Growth Operational Programme). TV was supported by the National Research, Development and Innovation Office NKFIH (Grant No. KH125096). K\.Z and AB are supported by NCN (Grant No. DEC-2015/18/A/ST2/00274).

\begin{widetext}

\begin{table*}[h]
\centering
\begin{tabular}{cccccccc} \hline \hline
~~$N$~~& ~~$k$~~& ~~purity~~   & ~~$W$~~ & ~~$M$~~& ~~~$F$~~~ & ~~~$f_{crit}$~~~& $p_v (\%)$\\ \hline
	2 & 1 & 1    &  0.5               &  $M_2 = 3$  &$F_x = 4$, $F_y =0$, $F_z = 4$& 0.293&28.32\\ \hline
	3 & 1 & 1    &  0.5               &  $M_3 = 4, M_2=3$ &$F_x=3, F_y=3, F_z=9$&0.5&74.69\\
	  & 2 & 1/4  & 0             &  $M_3 = 1$ &$F_x=0,F_y=3,F_z=3$&0&0\\ \hline
	4 & 1 & 1    & 0.5                & $M_4 = 9, M_2= 6$ &$F_x=4,F_y=4,F_z=16$&0.647&94.24\\
	  & 2 & 1/2  & 0.5                & $M_4 = 3, M_3 = 4$ &$F_x=F_y=F_z=4$&0.422&35.11\\
	  & 3 & 1/4  & 0                  & $M_4 = 3$&$F_x=F_y=F_z=4$&0.292& 0.024 \\ \hline
	5 & 1 & 1    & 0.5                & $M_5=16, M_4 = 5, M_3 = 10 $&$F_x=F_y=F_z=25$&0.75&99.60\\
    & 2 & 1    & 0.5                & $M_5 =6, M_4= 15, M_3 = 10 $&$F_x=F_y=F_z=5$&0.568& 99.96\\
	  & 3 & 1/2  & 0.5                & $M_4 = 15$ &$F_x=5, F_y=4, F_z=5$&0.460&63.65\\
	  & 4 & 1/16 &  0              & $M_5 = 1$&$F_x=0,F_y=5,F_z=5$&0&0 \\ \hline
	6 & 1 & 1    &  0.5       & $M_6=33, M_4=15, M_2=15$ &$F_x=6, F_y=6, F_z=36$&0.823&99.97\\
	  & 2 & 1    &  0.5    & $M_6=10, M_5=24, M_4=21, M_3=8$&$F_x=F_y=F_z=6$&0.666&$>99.99$\\
	  & 3 & 1    &  0.5      & $M_6=18, M_4=45$&$F_x=F_y=F_z=6$&0.591&100\\
	  & 4 & 1/16 &  0    & $M_6=1, M_5=2$ &$F_x=1, F_y=6, F_z=5$&0.293&$< 10^{-3}$\\
	  & 5 & 1/16 &  0        & $M_6=3$ &$F_x=F_y=F_z=6$&0.293&  $<10^{-6}$\\ \hline
	7 & 1 &  1   &  $>0$      & $M_7=64, M_6=7, M_4=35, M_2=21$  &$F_x=7, F_y=7, F_z=49$&0.875& 100\\
	  & 2 &  1   &  $>0$       & $M_7=15, M_6=42, M_5=42, M_4=21, M_3=7$ &$F_x=F_y=F_z=7$&0.785&~100\\
	  & 3 &  1/2 &  $>0$     & $M_6=42, M_4=21 $&$F_x=F_y=F_z=7$&0.644&99.16\\ \hline \hline
\end{tabular}
\caption{\label{tab} The properties of $k$-uniform $N$-qubit states given in Sec.~\ref{examp}: purity, $W$ - genuine multipartite entanglement monotone, $M$ - length of correlations, $F$ - quantum Fisher information, $F_i=F(\rho,J_i)$, where $J_i$ is the collective angular momentum operator, $f_{crit}$ - white noise robustness, and $p_v$ - probability of violation of local realism.}

\end{table*}

\end{widetext}

\appendix

\section{Numerical method based on nonlinear optimization}

\label{num}

The $k$-uniform states were found numerically by searching over the complete set of multipartite quantum states.
This procedure requires a non-linear optimization which was provided by Nlopt Package. We implemented PRAXIS (PRincipal AXIS) optimization routine which is an algorithm for gradient-free local optimization based on Richard Brent’s 'principal axis method' \cite{met}, specially designed for unconstrained optimization.

To determine the $k$-uniform states, we introduce a cost function defined in the following way,
\begin{equation}
p_{\max}(N,k) = \max_{\rho}\Bigg[ p_N^{k} - \beta \sum_{i=1}^k M_i(\rho) \Bigg],
\end{equation}
where $p_{\max}$ is the sum over the lengths of all non-zero correlations, maximized over the entire state space of $N$ parties. According to the definition of a $k$-uniform state, the correlations between the subsystems up to total of $k$ subsystems should vanish, whereas the rest is incorporated in the term $p_N^{k}$, the total length of the non-vanishing part of the correlations. To ensure that the constraint of vanishing correlations has been satisfied, we associate a regression coefficient $\beta$ to the lengths of correlations among the $k$ subsystems. To efficiently determine the global maximum for the cost function one takes the constant $\beta$ large enough, for which cost part vanishes, hence  $\beta >2^{N-1}$.

\section{Numerical method based on semidefinite programming}

\label{sdpnum}

To find $N$-party $k$-uniform states $\rho_N^k$ of a high purity ${\rm Tr}((\rho_N^k)^2)$, we use the following iterative procedure based on semidefinite programming. Inputs to the algorithm are the number of parties $N$, the number of subsystems $k$ with vanishing correlations and the dimension $d$ of local Hilbert spaces, which is assumed to be constant for all parties. In addition we fix the parameter $\epsilon\in[0,1]$, which sets the speed of convergence. Typical value of $\epsilon$ used in the algorithm is $0.3$.

Our task is to compute the optimal value
$P_{\rm{opt}}=\max_{\rho}{\,\rm Tr}(\rho^2)$
over $k$-uniform states $\rho\in \mathbb{C}^{(d^N)}$. In this problem the objective function is quadratic in the variable $\rho$ and the constraints are either semidefinite ($\rho\ge 0$) or linear ($k$-uniformity and normalization $\rm Tr(\rho)=1$). This is computationally a hard problem. However, let us notice that the optimal value $P_{\rm{opt}}$ is identical to
$\max_{\rho,\sigma}{\,\rm Tr}(\rho\cdot\sigma)$, where optimization is carried out over $k$-uniform states $\rho,\sigma\in \mathbb{C}^{(d^N)}$. Indeed, it can be shown that for any pair of $k$-uniform states $\rho$ and $\sigma$ the state $\rho_{av}=(\rho+\sigma)/2$ fulfills the relation ${\rm Tr}(\rho_{av}^2)\ge {\rm Tr}(\rho\cdot\sigma)$, which in turn entails the above alternative form for the optimal value $P_{\rm{opt}}$. We use this latter form to provide a sew-saw type heuristic method for computing $P_{\rm{opt}}$.

To this end, we choose randomly a $k$-uniform state $\rho$ and maximize ${\rm Tr}(\rho\cdot\sigma)$ over $k$-uniform $\sigma$ states. Then we fix $\sigma$ and optimize the same objective function over $k$-uniform $\rho$ states. Each two steps can be formulated as a semidefinite program, which we repeat again and again until convergence of ${\rm Tr}(\rho\cdot\sigma)$ is achieved.


Explicitely, the iterative algorithm described above looks as follows:

\begin{enumerate}
	
	\item Generate randomly a $k$-uniform state $\rho\in \mathbb{C}^{(d^N)}$.
	
	\item Solve the semidefinite program below.
	\begin{eqnarray}
	P = \max_{\sigma} && {\,\rm Tr}(\rho\cdot\sigma),\nonumber\\
	\textrm{s.t. }&& \rho_{\epsilon}\ge 0, {\rm Tr}(\rho_{\epsilon})=1, \nonumber \\
	&& \sigma = (1-\epsilon)\rho + \epsilon\rho_{\epsilon},\nonumber\\
	&&  \sigma \text{ is } k\text{-uniform},
	\label{eq:sdp}
	\end{eqnarray}
	where the optimization is carried out over the set of $k$-uniform density matrices $\sigma$, and the constraints within the optimization are either linear or semidefinite. For small enough $N$ and $d$ this problem can be solved efficiently.
	
	\item Set $\rho=\sigma$.
	
	\item Repeat steps 2-4 until convergence of the value $P$ is reached. $P$ defines a lower bound to the value of $P_{\rm{opt}}$.
	
\end{enumerate}

Note that it may not be easy to generate randomly $N$-party $k$-uniform states within step~1. We can sidestep this issue by generating instead a random $N$-party state and setting $\epsilon=1$ within the very first iteration. Then step 2 will ensure that $\sigma$ is $k$-uniform, hence $\rho$ in step 3 will also be $k$-uniform. Also notice that the value of $P$ is non-decreasing with the sequence of iterations. However, the above optimization may still get stuck in local maxima of the function ${\rm Tr}(\rho\cdot\sigma)$. Therefore, we may have to run the above procedure several times before obtaining a global optimal solution for $P_{\rm{opt}}$.

\end{document}